\documentclass{article}
\usepackage[T1]{fontenc}
\usepackage[latin1]{inputenc}
\usepackage{geometry}
\usepackage{graphics}
\usepackage{setspace}
\makeatletter

\providecommand{\LyX}{L\kern-.1667em\lower.25em\hbox{Y}\kern-.125emX\@}

 \newcommand{\lyxaddress}[1]{
   \par {\raggedright #1 
   \vspace{1.4em}
   \noindent\par}
 }

\makeatother

\begin{document}

\title{Persistence Exponents and Scaling In Two Dimensional XY model and A Nematic
Model }

\author{Subhrajit Dutta\thanks{
electronic address : subhro@juphys.ernet.in
} and Soumen Kumar Roy\thanks{
electronic address : skroy@juphys.ernet.in
}}

\maketitle

\lyxaddress{~~~~~~~~~~~~~~~~~~~~~~~~~~~~~~~~~Department
of Physics, Jadavpur University, Calcutta-700032, INDIA}

\begin{abstract}
{\large The persistence exponents associated with the T=0 quenching dynamics
of the two dimensional XY model and a two dimensional uniaxial spin nematic
model have been evaluated using a numerical simulation. The site persistence
or the probability that the sign of a local spin component does not change starting
from initial time t=0 up to certain time t, is found to decay as \( L(t)^{-\theta } \),
(L(t) is the linear domain length scale ), with \( \theta =0.305 \) for the
two dimensional XY model and 0.199 for the two dimensional uniaxial spin nematic
model. We have also investigated the scaling (at the late time of phase ordering)
associated with the correlated persistent sites in both models. The persistence
correlation length was found to grow in same way as \( L(t) \).} {\Large }{\Large \par}

{\Large 1. Introduction :}{\Large \par}
\end{abstract}
{\large Phase ordering of various systems with scalar, vector and more complex
order parameters has been an active field of research over last few years\cite{brayrev, furu}.
When a system is suddenly quenched from a high temperature homogeneous equilibrium
phase in to an ordered phase ( at temperature less than the critical temperature,
\( T_{c} \) ), the system does not get ordered suddenly. Instead domains of
various degenerate phases grow and in the thermodynamic limit the system develops
a length scale that grows with time without any upper bound. Recently we have
studied the coarsening dynamics of the two dimensional quenched uniaxial nematic
\cite{dutta}, where it has been established using a cell dynamic scheme \cite{puri},
that in a zero temperature quenched two dimensional nematic lattice model, dynamical
scaling is obeyed and the growth law associated with the linear length scale
of domains (L(t)) is similar to that in the two dimensional XY model \cite{rojas}.
In both the systems, asymptotically, the domain length scale \( L(t) \), was
found to grow as \( (t/lnt)^{1/2}. \) Although the interaction Hamiltonians
have different symmetry, the similar structure of the topological defects supported
by these models \cite{dutta, enersc} (both models possess stable point topological
singularity), is responsible for similar asymptotic growth law of \( L(t) \).
So from the point of view of growth law associated with the dynamical domain
length scale \( L(t) \), the coarsening dynamics are indistinguishable. However
when we look for more detailed correlation that exists within the dynamically
evolving non-equilibrium system, it may be possible that the two models will
show different features. One such physical quantity, which probes the details
of the history of the dynamics, is the persistence probability or simply the
persistence. Persistence is an interesting property from both theoretical and
experimental point of view in the field of non-equilibrium statistical mechanics
\cite{watson}.}{\large \par}

\begin{figure}
{\par\centering \resizebox*{0.6\textwidth}{!}{\includegraphics{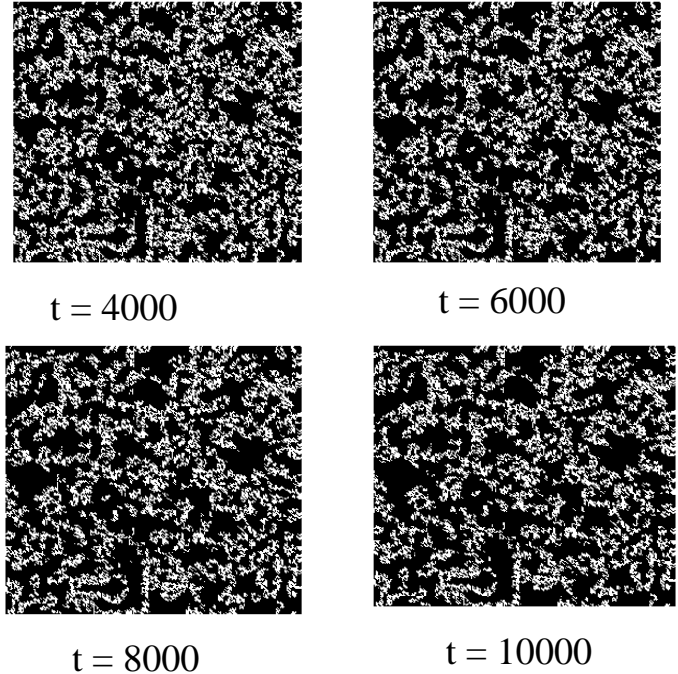}} \par}

\caption{{\large The persistent spins in 200 x 200 2d XY model for t=4000, 6000, 8000
and 10000 after the system is quenched from a high temperature initial stage
to T=0 (white portions represent persistent sites)}.}
\end{figure}

\begin{figure}
{\par\centering \resizebox*{0.6\textwidth}{!}{\includegraphics{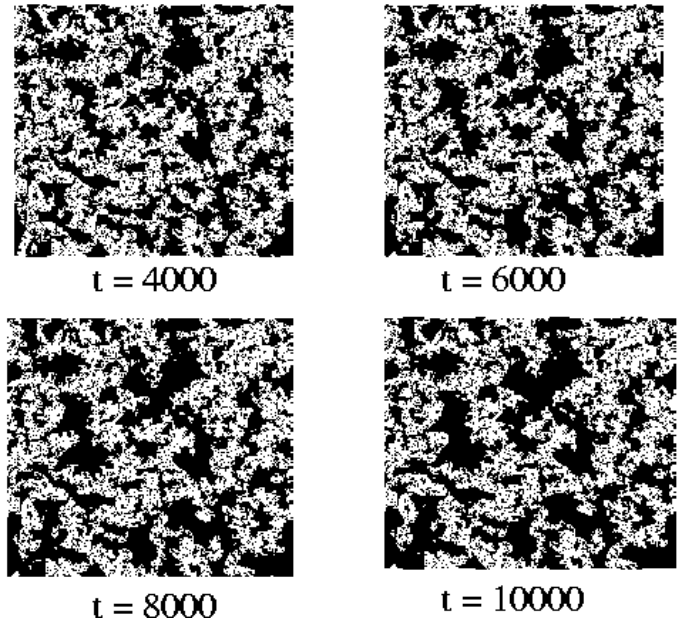}} \par}

\caption{{\large The persistent spins in 200 x 200 2d spin nematic model for t=4000,
6000, 8000 and 10000 after the system is quenched from a high temperature initial
stage to T=0 (white portions represent persistent sites).}\large }
\end{figure}

{\large Persistence in a general non-equilibrium process may be defined as the
probability that any zero-mean stochastic variable \( X(t) \), does not change
sign up to certain time t starting from an initial time t=0. Study of persistence
in various non-equilibrium systems is of recent interest \cite{snrev, purrev}.
Here one studies the time evolution of the order parameter field, \( \phi (x,t) \),
which varies in space as well as in time. Persistence in a general extended
non-equilibrium system may be defined as the probability that some local order
parameter (fixed at a particular point x in space) has not changed sign up to
a certain time t starting from the initial time t=0. This is more properly called
the local or site persistence (another quantity, which is of relevant interest
in study of non-equilibrium systems, is global persistence \cite{global1, global2},
which is defined in the same way for the total value of the order parameter).
The definition of persistence justifies that, it probes into the history of
the evolution, so the analytical calculation is difficult. The difficulty associated
with the calculation of persistence in case of a general non-equilibrium process
may be understood if we look for the two time correlator of the process with
the strong restriction that the process is Gaussian and a stationary one {[}12\( - \)14{]}.
A stationary Gaussian process, is completely determined by its two time correlator
\( C(\tau )=<X(t)X(t+\tau )> \). However the persistence probability is known
exactly only for few correlators. For the Stationary Gaussian Markovian correlator
\( C(\tau )=e^{-a\tau } \), the persistence probability is given by, \( p(\tau )= \)\( 2/\pi  \)
\( arcsin(e^{-a\tau }) \) and in large time it decays as \( 2/\pi  \) \( e^{-a\tau } \)
. But the Gaussian Stationary non-Markovian correlators can not expressed in
pure exponential form and the persistence probability sensitively depends on
the full functional form of the correlator, not just on its form in the asymptotic
limit of time. Hence, the decay of persistence for non-Markovian processes is
a nontrivial one and can not be determined exactly. Thus nontrivial decay of
persistence simply reflects the non-Markovianness associated with the process.
For the simple random walk problem, which is a Markovian process, trivial decay
of persistence is observed \cite{snrev}. In a general non-equilibrium dynamics,
the normalized two time correlator in asymptotic limit of time may be written
in the simple scaling form \( f(L(t)/L(t^{'})) \) ( with \( t<t^{'} \) and
with the assumed validity of dynamical scaling), where \( L(t) \) is the diverging
dynamic length scale associated with the domains in a coarsening system. Clearly
this process is nonstationary in real time. However if one makes the transformation
\( u=lnL(t) \), then the evolution of the normalized stochastic process \( (X(t)/\sqrt{<X(t)^{2}>}) \)
becomes stationary in the logarithmic scale u and the persistence probability
for a Gaussian process in asymptotic limit decays as \( e^{-\theta u} \) or
simply as \( L(t)^{-\theta } \) \cite{block1}, where \( \theta  \) is known
as persistence exponent. In some of the papers on persistence, P(t) is assumed
to decay as \( t^{-\theta ^{'}} \), although in general it should decay as
\( L(t)^{-\theta } \). This is because \( L(t)\sim t \)\( ^{1/z} \) is not
always true, z being the dynamic growth exponent associated with the growth
law of L(t), (e.g. in the present systems \( L(t)\sim (t/lnt)^{1/2}) \) , and
hence P(t) is not always of the form \( t^{-\theta ^{'}} \) \cite{private}.
So it will be more appropriate to designate the power of \( L(t) \) in the
decay as the persistence exponent.}{\large \par}

{\large The exponent \( \theta  \) comes out to be independent of other dynamic
exponents like the dynamic growth exponent \( z \) and autocorrelation or Fisher-Huse
exponent \( \lambda  \) \cite{brayrev} ( in the scaling regime the two time
correlation function or the autocorrelation function is given by, \( C(t,t^{'})\sim  \)\( (L(t)/L(t^{'}))^{\lambda } \),
for \( t^{'}>>t \)). As said earlier, the nontriviality associated with the
persistence exponent, simply reflects the non-Markovianness associated with
the process. For the case of simple 1-d random walk problem, which is a Markovian
one, the persistence probability is found to decay as \( p(t)\sim  \)\( t^{-\theta } \)
with \( \theta  \) exactly equal to 1/2 \cite{snrev, feller}. However most
of the noneqilibrium dynamical processes are non-Markovian and hence nontrivial
decay of persistence is generally observed. For example, even in the simple
scalar diffusion equation (where the stationary two time correlator \( C(\tau ) \) is
equal to \( [sech(\tau /2)]^{d/2} \)\cite{diffusion1}, which is significantly
different from the pure exponential form and hence non-Markovian ) with random
initial conditions, the nontrivial algebraic decay \( (t^{-\theta }) \) of
P(t) has been observed and analytically \( \theta  \) was calculated using
an Independent Interval Approximation (IIA). In the IIA, the interval of time
{[}0,t{]} is divided into independent zero crossing (where the zero mean stochastic
variable changes sign) intervals or in the language of probability, the distribution
of successive zero crossing intervals written as the product of their individual
distribution \cite{diffusion1}. But IIA estimates could not be systematically
improved, for which series expansion methods were used \cite{diffusion2}. The
IIA estimates of \( \theta  \) were, 0.1203 for d=1, 0.1862 for d=2 and 0.2358
for d=3 diffusion models and these values are in good agreement with the simulation
results\cite{diffusion1}. Independent Interval Approximation could not be applied
to those systems where the zeros are not uniformly distributed over a time interval
\cite{snrev}. }{\large \par}

{\large A simple example of coarsening system is the zero temperature Glauber
dynamics of 1-d Potts model. Even in this simple one dimensional system the
persistence exponent is nontrivial \cite{1dpotts} and Derrida et. al. could
give an exact solution \cite{1dpotts2}. However the technique used could not
be extended to higher experimentally relevant dimensions. For 1-d Potts case,
the value of persistence exponent \( \theta (q) \) (q is the Potts state) was
found exactly to be \( - \)1/8+2/\( \pi ^{2} \){[}arccos(2\( - \)q/\( \sqrt{2} \)q){]}\( ^{2} \)
\cite{1dpotts2}. So for 1-d Ising model \( P(t) \) was found to decay as \( t^{-3/8} \)
or \( L(t)^{-3/4} \) (since \( L(t)\sim  \) \( t^{1/2} \), in case of 1-d
Potts case). Calculation of persistence for Glauber Ising case can not be done
using IIA, because the concerned process is non-smooth (where the zeroes are
not uniformly distributed). The exponents were estimated using an approximate
method by Majumdar and Sire \cite{GCA} based on the frame work of Gaussian
Closure Approximation {[}GCA{]} (in GCA, the Ising spins are assumed to be the
sign of a Gaussian function), using variational approach by choosing the Hamiltonian
of a quantum harmonic oscillator as trial Hamiltonian. The exponents found by
this technique for Glauber Ising cases in d=1,2 and 3, were also confirmed numerically
\cite{GCA}. The exponent \( \theta  \) of 2d Ising ( 0.195 using GCA ) model,
was confirmed experimentally using a twisted nematic film which effectively
coarsens via Glauber dynamics \cite{yurke}. The persistence exponent depends
on the updating rule of the dynamics. In T=0 dynamics of 1-d Potts case the
value of \( \theta  \) using parallel (synchronous) updating rule was found
to be exactly double as that of serial (asynchronous) updating rule \cite{purrev, menon}.
Study related to finite temperature persistence \cite{block1,replica,stauffer, block2},
reveals that the temperature universality is not broken by this new exponent\cite{ block1, stauffer}.}{\large \par}

{\large In the present work we have performed a numerical simulation to obtain
the persistence exponent associated with the T=0 quenching dynamics of the two
dimensional spin models. These are the XY model and the uniaxial spin nematic
model (where the spin dimensionality is three). As already stated, both systems
obey dynamical scaing in a T=0 quench and the domain length scales as \( (t/lnt)^{1/2} \)
in the asymptotic limit. The purpose of the present study is two fold. It is,
to our knowledge, the only work so far on the study of persistence exponent
in a continuous spin system and secondly we have investigated if the persistence
exponents differ in the two systems which exhibit the same asymptotic dynamical
scaling growth law. We have also investigated the scaling associated with the
correlated persistence sites in both models and these were found to grow in
the same way as L(t).}

\begin{figure}
{\par\centering \resizebox*{0.8\textwidth}{!}{\includegraphics{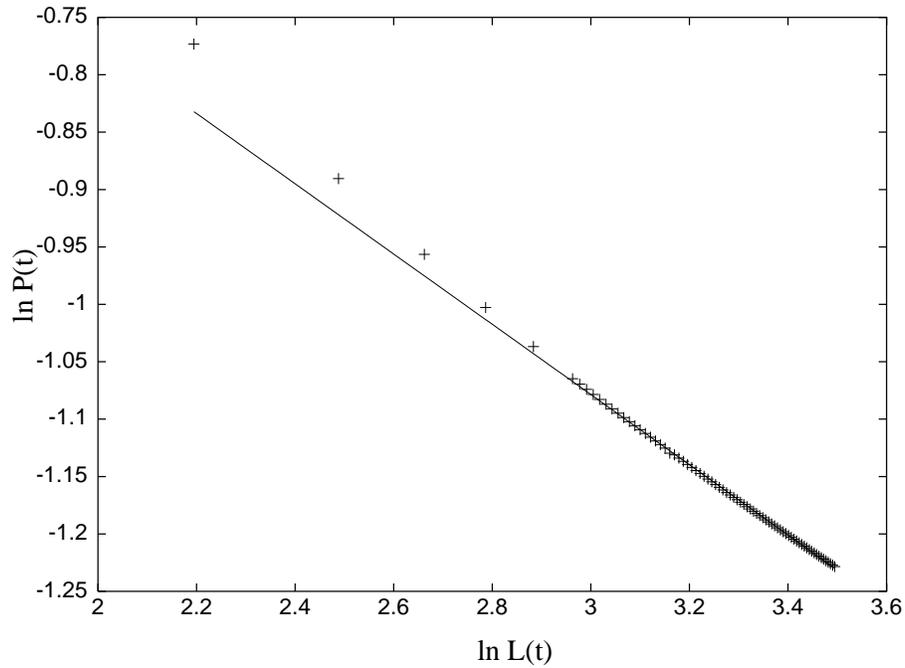}} \par}

\caption{{\large Plot of lnP(t) against ln (L(t)) for 400 x 400 XY model. The linearity
of the plot in the asymptotic time limit ensures the decay of the form P(t)=
L(t)\protect\( ^{-\theta }\protect \)or (t/lnt)\protect\( ^{-\theta /2}\protect \),
with \protect\( \theta =0.305\protect \). The linear region extends from t=3000
to t=10000. Average over 12 initial configurations and 400x400 sites were taken. }\large }
\end{figure}
{\Large ~~~~~~~~}

{\Large 2. Simulation Techniques : }{\Large \par}

{\large The Hamiltonian of two-dimensional XY model is given by,}{\large \par}

\textbf{\[
H=-\sum _{<i,j>}(\phi _{i},\phi _{j})\]
}{\large Where \( \phi  \) is usual two dimensional vector spin and <i,j> represents
nearest neighbor sites. The equation of motion is given by \cite{newman}, }{\large \par}

\[
\frac{\partial \phi _{i}}{\partial t}=\sum _{j}\phi _{j}-\sum _{j}(\phi _{i},\phi _{j})\phi _{j}\]
{\large where sum is taken over nearest neighbor sites. We have omitted any
noise in the equation of motion, hence we are effectively working at T=0 . }{\large \par}

{\large The Hamiltonian of the two-dimensional model representing the uniaxial
nematic, is given by ,}{\large \par}

\[
H=-\sum _{<i,j>}(\phi _{i},\phi _{j})^{2}\]

{\large where \( \phi  \) is the usual three dimensional vector spin on a two
dimensional lattice. In this model in addition to O(3) symmetry, there exists
local inversion symmetry and hence it represents a uniaxial nematic. It is also
known as the spin nematic model and resembles the celebrated Lebwohl-Lasher
model for uniaxial nematic, where the nearest neighbor interaction is proportional
to \( -P_{2}(cos\theta ) \) (\( P_{2} \) is the second Legendre polynomial
and \( \theta  \) is the angle between two nearest neighbor spin vectors) \cite{lebwohl}.
Similar to two dimensional XY case, the equation of motion is given by \cite{blundell}, }{\large \par}

\[
\frac{\partial \phi _{i}}{\partial t}=\sum _{j}(\phi _{i},\phi _{j})\phi _{j}-\sum _{j}(\phi _{i},\phi _{j})^{2}\phi _{j}\]

{\large where the sum is taken over nearest neighbor sites.}{\large \par}

{\large We have performed numerical simulation of discretized versions of the
equations of motion. The time step \( \delta t \) was taken to be 0.02. However
all the results shown in this paper, were found to be independent of \( \delta t \)
(for \( \delta t \) <0.1) in the asymptotic regime. We have presented here
results for a 400 x 400 lattice. We did not observe any significant finite size
effect by comparing the results obtained for smaller lattice sizes. }{\large \par}

~~~~~~

{\Large 3. Persistence Probability and Scaling of Persistence Correlation :}{\Large \par}

{\large The persistence probability P(t) for continuous spin system may be defined
as the probability that starting from the initial time t=0, any one of the components
of the continuous spin at a fixed position in the lattice does not change its
sign up to time t. Owing to the symmetry, one must average it over all components.
We have taken the average over several random initial configurations as well
as over all lattice sites. Mathematically we can write the persistence probability
as, }{\large \par}

{\large \[
P(t)=Probability[S_{i}(t^{'})\times S_{i}(0)>0,\, for\, all\, t^{'}\, in\, [0,t]]\]
} {\large Where S\( _{i} \) is the i\( ^{th} \)component of the spin vector
at a particular lattice site. }{\large \par}

\begin{figure}
{\par\centering \resizebox*{0.8\textwidth}{!}{\includegraphics{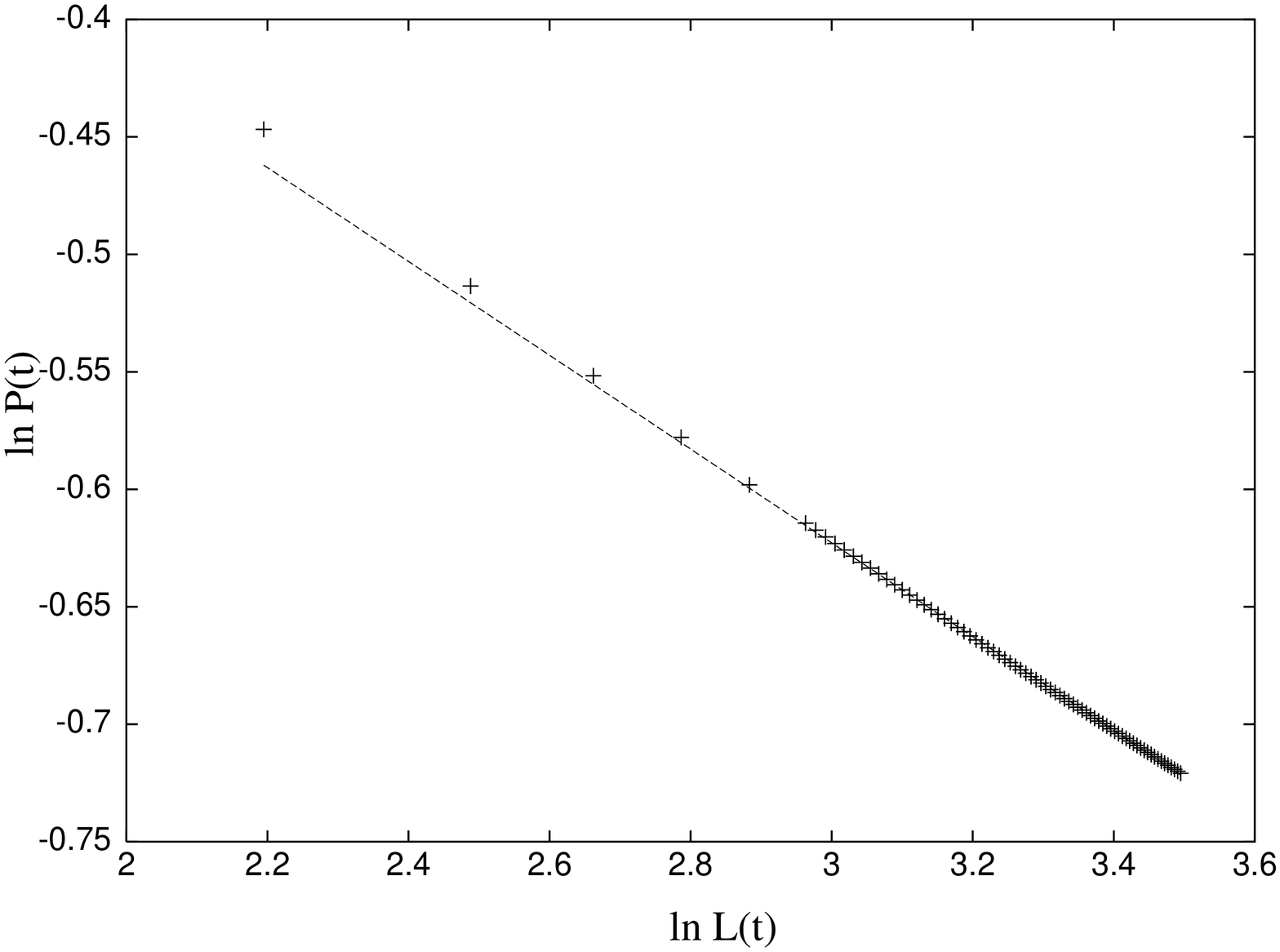}} \par}

\caption{{\large Plot of lnP(t) against lnL(t) for 400x400 spin nematic model. The linearity
of the plot in the asymptotic limit ensures the decay P(t)= L(t)\protect\( ^{-\theta }\protect \)or
(t/lnt)\protect\( ^{-\theta /2}\protect \), with \protect\( \theta \protect \)
=0.199. The linear region extends from t=3000 to t=10000. Average over 15 initial
configurations and 400x400 sites were taken. }\large }
\end{figure}
 {\large Scaling and fractal formation of the correlated persistence sites have
achieved recent interest by various researchers \cite{manoj, jain,braym3}.
In the present work we have investigated scaling in the spatial correlation
of the persistence sites. For this we have evaluated the normalized two point
corrector,}

\[
C(r,t)=<n_{i}(t)n_{i+r}(t)>/<n_{i}(t)>\]

{\large where, < > represents the average over sites as well as random initial
conditions. n\( _{i} \)(t)=1 if the i\( ^{th} \) site is persistent otherwise
it is 0. This correlation just gives the probability that the spin at \( (i+r)^{th} \)
site is persistent, given the i\( ^{th} \) site is persistent. Beyond a certain
length \( \xi (t) \) (persistence correlation length) the sites are found to
be uncorrelated and C(r,t) is simply \( <n(t)> \) or the persistence probability
P(t). However for \( r<\xi (t) \) there exists strong correlation. In the correlated
region \( C(r,t) \) shows a power law decay with distance \( r^{-\alpha } \)
and hence is independent of t or L(t). So for \( r<\xi (t) \), there exists
strong correlation with scale invariant behavior, which indicates the expected
self similar fractal structure formed by the persistent sites \cite{purrev,manoj, jain}.
Now at \( r=\xi (t) \), consistency demands, \( \xi ^{-\alpha }(t)=L(t)^{-\theta } \)
(since \( P(t) \)\( \sim  \)\( L(t)^{-\theta } \)), which simply implies
\( \xi (t) \) should diverge as \( L(t)^{\zeta } \) with \( \zeta =\theta /\alpha  \).
Mathematically we can write C(r,t) as,}{\large \par}

{\large \begin{eqnarray*}
C(r,t) & =\, r^{-\alpha } & for\, r\, \ll \xi (t)\\
 & P(t) & for\, r\, \gg \xi (t)
\end{eqnarray*}
} 

{\large Clearly in scaling form C(r,t) can be written as, }{\large \par}

{\large \[
C(r,t)=P(t)f(r/\xi (t))\]
}{\large \par}

\begin{figure}
{\par\centering \resizebox*{0.8\textwidth}{!}{\includegraphics{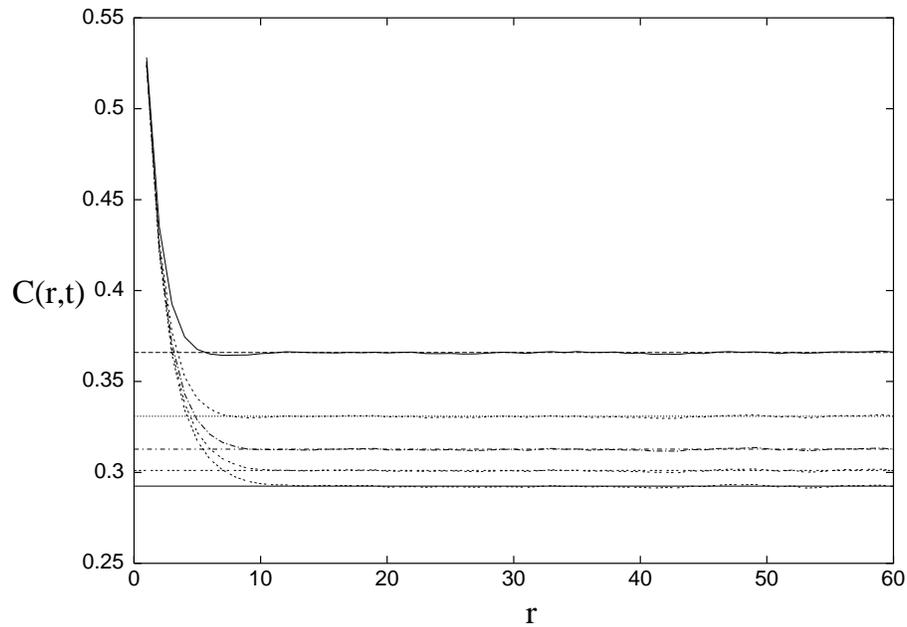}} \par}

\caption{{\large The variation of correlation function with distance for the 400 x 400
2d XY model. For small values of r, C(r,t) is independent of t. For large r,
it is same as persistence probability (lines parallel to x -axis represents
P(t)). The data are for time steps t= 2000, 4000, 6000, 8000 and 10000 (from
top to bottom) with persistence probability P(t) = 0.366, 0.331, 0.313, 0.301
and 0.292 respectively.}\large }
\end{figure}
\break
\begin{figure}
{\par\centering \resizebox*{0.8\textwidth}{!}{\includegraphics{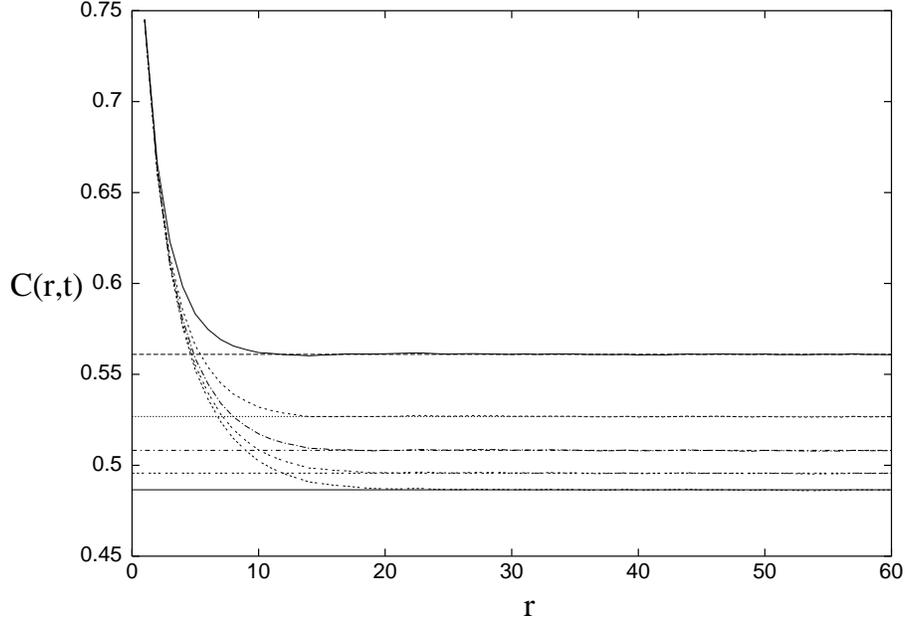}} \par}

\caption{{\large The variation of correlation function with distance for the 400 x 400
2d spin nematic model. For small values of r, C(r,t) is independent of t. For
large r, it is same as persistence probability (lines parallel to x-axis represents
P(t)). The data are for time steps t= 2000, 4000, 6000, 8000 and 10000 (from
top to bottom) with persistence probability P(t) = 0.561, 0.527, 0.508, 0.496
and 0.487 respectively.}\large }
\end{figure}

{\large Where f(x) is given by, }{\large \par}

{\large \begin{eqnarray*}
f(x)= & x^{-\alpha } & for\, x\, \ll 1\\
 & 1 & for\, x\, \gg 1
\end{eqnarray*}
}{\large \par}

{\Large 4. Results and Discussions :}{\Large \par}

{\large In Figure 1 and Figure 2 we have shown how correlated regions of persistence
sites are formed in the two dimensional XY and the two dimensional spin nematic
models at various times t, after the system was quenched from the initial homogeneous
T=\( \infty  \) configuration. In Figure 3 we have shown the decay of the persistence
with L(t)=(t/lnt)\( ^{1/2} \) for the two dimensional XY model. The linearity
in the log-log plot reflects the decay to be of the form, \( P(t)=L(t)^{-\theta } \)
or \( (t/lnt)^{-\theta /2} \) in the late time regime. The exponent \( \theta  \)
we obtained was 0.305. In figure 4 we have depicted the same for two dimensional
spin nematic model and the exponent \( \theta  \) we obtained was 0.199. }{\large \par}

{\large In Figure 5 and Figure 6 we have shown the correlator C(r,t) plotted
against r for various values of t for XY model and the spin nematic model. In
both figures it is observed that for small value of r, C(r,t) for each time
overlaps and for large values of r, C(r,t) is equal to P(t). For small value
of r, a r\( ^{-\alpha } \) decay is observed. In figure 7 and 8 we have shown
the log-log plot of scaling function of C(r,t ) for the XY model and the spin
nematic model. We obtained good collapse for \( \zeta =1 \) (and hence \( \alpha =\theta ) \)
which implies that the persistence correlation length \( \xi (t) \) diverges
as L(t) or (t/lnt)\( ^{1/2} \) . It is of interest to note that the persistence
correlation length has similar divergence as that of the length scale associated
with the domains formed during coarsening of the system. We point out that,
we have also tried to collapse our data with the familiar form of the growth
law t\( ^{1/2} \), but were unable to obtain a good collapse. }{\large \par}

\begin{figure}
{\par\centering \resizebox*{0.8\textwidth}{!}{\includegraphics{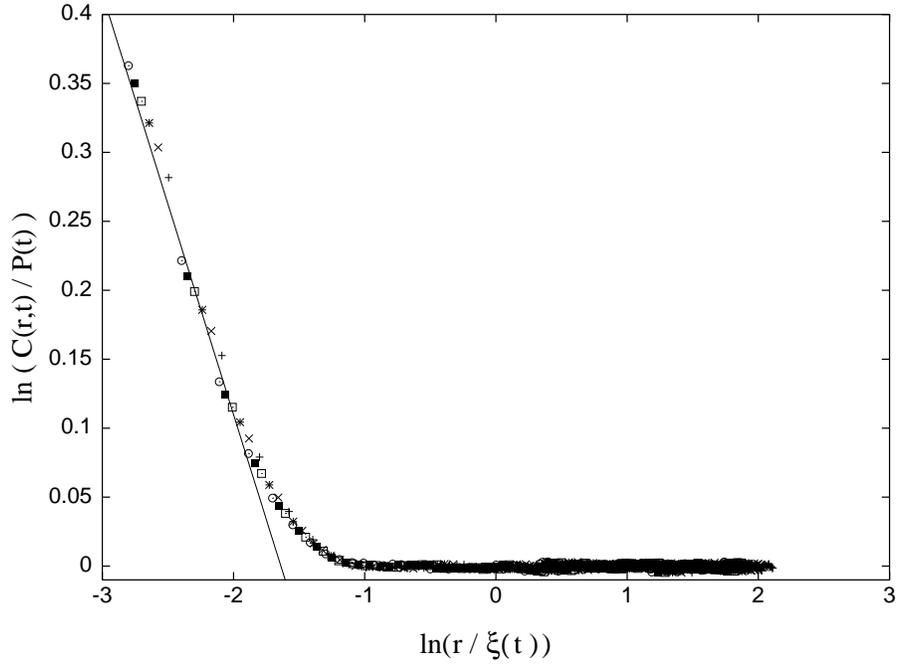}} \par}

\caption{{\large Plot of ln(C(r,t)/P(t)) against ln(r\protect\( /\xi (t)\protect \)).
The best collapse is obtained when the value of \protect\( \zeta =1\protect \),
i.e. if \protect\( \xi (t)\protect \) \protect\( \sim \protect \) (t/lnt)\protect\( ^{1/2}\protect \).
The straight line for small values of \protect\( r/\xi (t)\protect \), has
slope \protect\( \alpha \protect \) equal to 0.305. Which is equal to the persistence
exponent of the 2d XY model. The data used are for time t=5000, 6000, 7000,
8000, 9000 and 10000.}\large }
\end{figure}

\begin{figure}
{\par\centering \resizebox*{0.8\textwidth}{!}{\includegraphics{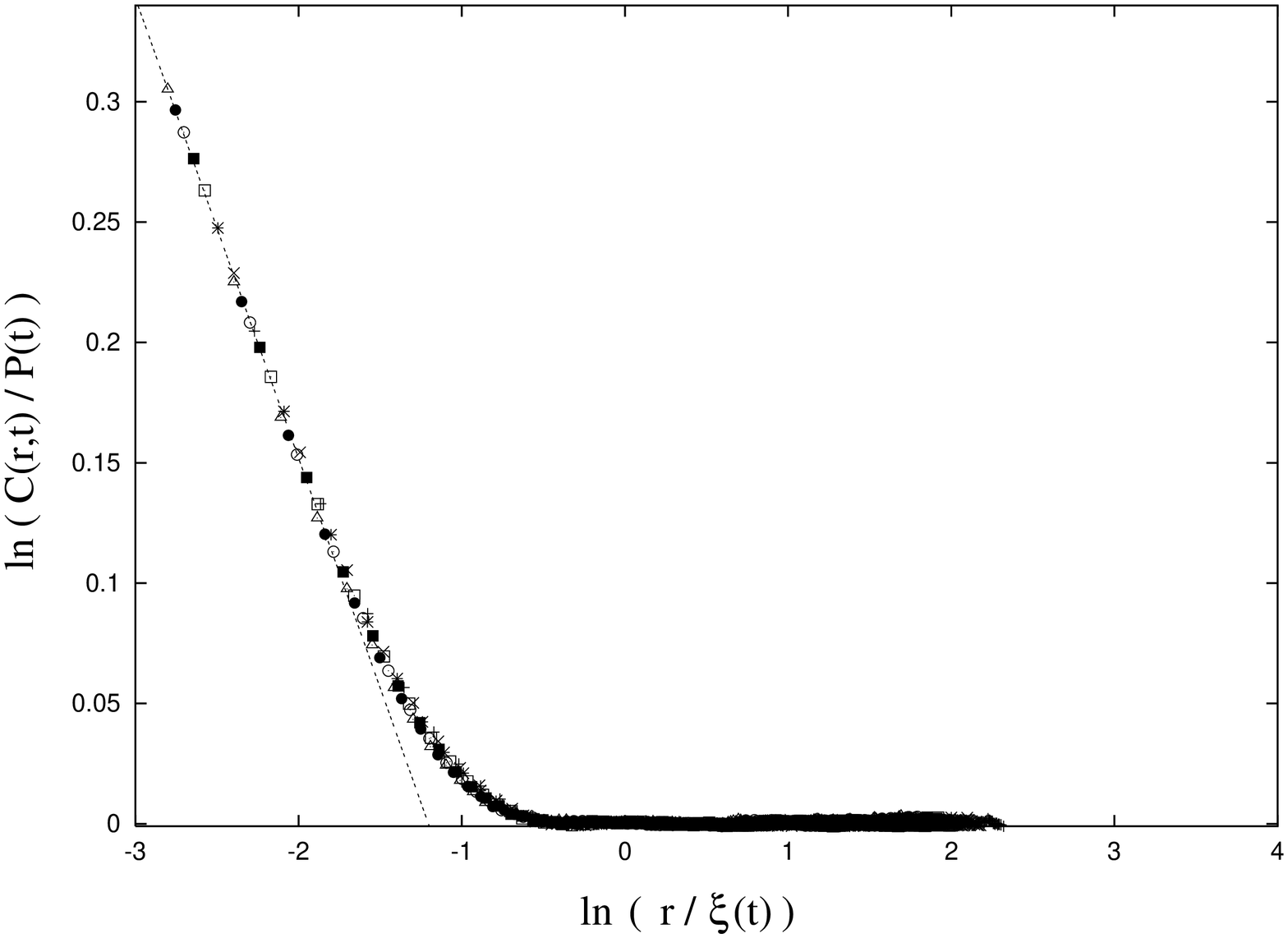}} \par}

\caption{{\large Plot of ln(C(r,t)/P(t)) against ln(r/\protect\( \xi (t)\protect \)
). The best collapse is obtained when the value of \protect\( \zeta =1\protect \),
i.e. if \protect\( \xi (t)\protect \) \protect\( \sim \protect \) (t/lnt)\protect\( ^{1/2}\protect \).
The straight line for small values of \protect\( r/\xi (t)\protect \), has
slope \protect\( \alpha \protect \) equal to 0.191, which is almost equal to
the persistence exponent of 2d spin nematic model. The data used are for time
t=3000, 4000, 5000, 6000, 7000, 8000, 9000 and 10000.}\large }
\end{figure}

{\Large 5. Conclusion}{\Large \par}

{\large We would like to summarize the main findings of the paper. In the present
work we have studied the site persistence in the T=0 quenching dynamics of the
two dimensional XY model and two dimensional spin nematic model. Although in
both the models, the dynamical domain length scales L(t), have similar growth
laws in asymptotic limit, the persistence exponents comes out to be different.
In the XY model, it is 0.305 while in the spin nematic model it is 0.199. We
have also investigated the scaling structure of persistence sites for both the
models. We got the growth law of persistence correlation length to be the same
as that of the domain length scale L(t), i.e. \( (t/lnt)^{1/2} \). }{\large \par}

{\large Acknowledgments : The authors would like to thank Prof. Clement Sire
and Prof. Purusattam Ray for some useful discussions. We also thank Prof. A.
J. Bray for bringing ref. 18 and 32 to our notice. One of us S. D. acknowledges
financial support from Council of Scientific and Industrial Research, India.}{\large \par}

\break

\textbf{LIST OF FIGURE CAPTIONS :}

{\large figure 1 : The persistent spins in 200 x 200 2d XY model for t=4000,
6000, 8000 and 10000 after the system is quenched from a high temperature initial
stage to T=0 (white portions represent persistent sites).}{\large \par}

{\large figure 2: The persistent spins in 200 x 200 2d spin nematic model for
t=4000, 6000, 8000 and 10000 after the system is quenched from a high temperature
initial stage to T=0 (white portions represent persistent sites).}{\large \par}

{\large figure 3 : Plot of lnP(t) against ln (L(t)) for 400 x 400 XY model.
The linearity of the plot in the asymptotic time limit ensures the decay of
the form P(t)= L(t)\( ^{-\theta } \)or (t/lnt)\( ^{-\theta /2} \), with \( \theta =0.305 \).
The linear region extends from t=3000 to t=10000. Average over 12 initial configurations
and 400x400 sites were taken.}{\large \par}

{\large figure 4 : Plot of lnP(t) against lnL(t) for 400x400 spin nematic model.
The linearity of the plot in the asymptotic limit ensures the decay P(t)= L(t)\( ^{-\theta } \)or
(t/lnt)\( ^{-\theta /2} \), with \( \theta  \) =0.199. The linear region extends
from t=3000 to t=10000. Average over 15 initial configurations and 400x400 sites
were taken. }{\large \par}

{\large figure 5 : The variation of correlation function with distance for the
400 x 400 2d XY model. For small values of r, C(r,t) is independent of t. For
large r, it is same as persistence probability (lines parallel to x -axis represents
P(t)). The data are for time steps t= 2000, 4000, 6000, 8000 and 10000 (from
top to bottom) with persistence probability P(t) = 0.366, 0.331, 0.313, 0.301
and 0.292 respectively.}{\large \par}

{\large figure 6 : The variation of correlation function with distance for the
400 x 400 2d spin nematic model. For small values of r, C(r,t) is independent
of t. For large r, it is same as persistence probability (lines parallel to
x-axis represents P(t)). The data are for time steps t= 2000, 4000, 6000, 8000
and 10000 (from top to bottom) with persistence probability P(t) = 0.561, 0.527,
0.508, 0.496 and 0.487 respectively.}{\large \par}

{\large figure 7 : Plot of ln(C(r,t)/P(t)) against ln(r\( /\xi (t) \)). The
best collapse is obtained when the value of \( \zeta =1 \), i.e. if \( \xi (t) \)
\( \sim  \) (t/lnt)\( ^{1/2} \). The straight line for small values of \( r/\xi (t) \),
has slope \( \alpha  \) equal to 0.305. Which is equal to the persistence exponent
of the 2d XY model. The data used are for time t=5000, 6000, 7000, 8000, 9000
and 10000.}{\large \par}

{\large figure 8 : Plot of ln(C(r,t)/P(t)) against ln(r/\( \xi (t) \) ). The
best collapse is obtained when the value of \( \zeta =1 \), i.e. if \( \xi (t) \)
\( \sim  \) (t/lnt)\( ^{1/2} \). The straight line for small values of \( r/\xi (t) \),
has slope \( \alpha  \) equal to 0.191, which is almost equal to the persistence
exponent of 2d spin nematic model. The data used are for time t=3000, 4000,
5000, 6000, 7000, 8000, 9000 and 10000.}{\large \par}
\end{document}